\documentclass[paper]{ieice}
\usepackage[dvips]{graphicx}
\usepackage[fleqn]{amsmath}
\usepackage{newtxtext}
\usepackage[varg]{newtxmath}

\title{A portable and Linux capable RISC-V computer system in Verilog HDL}
\authorlist{%
 \authorentry[miura@arch.cs.titech.ac.jp]{Junya MIURA}{n}{TokyoTech}\MembershipNumber{}
 \authorentry[miyazaki@arch.cs.titech.ac.jp]{Hiromu MIYAZAKI}{s}{TokyoTech}\MembershipNumber{1900691}
 \authorentry[kise@c.titech.ac.jp]{Kenji KISE}{m}{TokyoTech}\MembershipNumber{0305928}
}
\affiliate[TokyoTech]{The authors are with the School of Computing, Tokyo Institute of Technology
}
\newcommand{\RVSoC}{RVSoC}
\newcommand{\RVCoreM}{RVCoreM}
\newcommand{\RVver}{v0.4.3}
\newcommand{\RVuc}{RVuc}

\begin{document}
\maketitle
%%%%%%%%%%%%%%%%%%%%%%%%%%%%%%%%%%%%%%%%%%%%%%%%%%%%%%%%%%%%%%%%%%%%%%%%%%%%%%%%%%%%%%%%%%%%%%%%%%%%
\begin{summary}
RISC-V is an open and royalty free instruction set architecture 
which has been developed at the University of California, Berkeley.
The processors using RISC-V can be designed and released freely.
Because of this, various processor cores and system on chips (SoCs) 
have been released so far.
However, there are a few public RISC-V computer systems that are portable and 
can boot Linux operating systems.

In this paper, we describe a portable and Linux capable RISC-V computer system 
targeting FPGAs in Verilog HDL.
This system can be implemented on an FPGA with fewer hardware resources, 
and can be implemented on low cost FPGAs or customized by introducing an accelerator.
This paper also describes the knowledge obtained through the development of 
this RISC-V computer system.
\end{summary}
%%%%%%%%%%%%%%%%%%%%%%%%%%%%%%%%%%%%%%%%%%%%%%%%%%%%%%%%%%%%%%%%%%%%%%%%%%%%%%%%%%%%%%%%%%%%%%%%%%%%


%%%%%%%%%%%%%%%%%%%%%%%%%%%%%%%%%%%%%%%%%%%%%%%%%%%%%%%%%%%%%%%%%%%%%%%%%%%%%%%%%%%%%%%%%%%%%%%%%%%%
\section{Introduction}

RISC-V~\cite{RISCV} is an emerging instruction set architecture (ISA) 
which has been developed at the University of California, Berkeley.
Because it is open and royalty free, 
processor cores or system on chips (SoCs) of RISC-V can be designed and released freely.
For this reason, 
processor cores and SoCs that operate on Field-Programmable Gate Arrays (FPGAs) 
such as Rocket chip~\cite{Rocket-chip} and BOOM~\cite{BOOM} have been released.
In addition, 
various software including some compilers and 
operating systems (OSs) begin to support the ISA. 

Linux is also built for RISC-V, on which many applications can run.
However, there are still a few RISC-V computer systems that support Linux.
Although Freedom U500 VC707 Dev Kit~\cite{freedom} from SiFive is available, 
it is difficult to use the system easily 
because an expensive FPGA is required.
Therefore, 
by developing a portable computer system of RISC-V that can be easily used 
on a small FPGA and support Linux, 
it is expected to be used in various fields.

In RISC-V~\cite{RISCV-SPEC},
many extended instruction sets are defined which can be added to
the base integer instruction set named {\it RV32I} that supports 32-bit address space.
The designers may select the supported instruction sets depending on the way to use the systems.
For example, the {\it M} extension for integer multiplication and division,
the {\it C} extension for the compressed instructions, and
the {\it F} extension for the single-precision floating-point instructions
are defined.
The notation of {\it RV32IM} indicates the ISA supporting RV32I and the M extension.
In order to support Linux on a RISC-V system, not only supporting RV32IM
but also supporting {\it RV32IMA} with the {\it A} extension for the atomic instructions
is necessary.

We design a portable and Linux capable RISC-V computer system supporting
{\it RV32IMAC} and implement it targetting a small FPGA in Verilog HDL.
This paper also describes the knowledge obtained through the development of this system.

%% The main contributions are as follows.
%% {\color{red}
%% \begin{itemize}
%% \item The changes required for Linux to operate from a simple RISC-V computer system are clarified.
%% \item We propose a new portable and Linux capable RISC-V computer system.
%% We show that the proposed system can be implemented on an 
%% FPGA board and various applications can be executed.
%% \item We evaluate the proposed RISC-V computer system and show its usefulness.
%% \end{itemize}
%% 
%% \vspace{10mm}
%% }

%%%%%%%%%%%%%%%%%%%%%%%%%%%%%%%%%%%%%%%%%%%%%%%%%%%%%%%%%%%%%%%%%%%%%%%%%%%%%%%%%%%%%%%%%%%%%%%%%%%%
\section{Challenges to support Linux}

We discuss the challenges to support Linux for
a baseline computer system like an embedded system which 
often used to control some simple hardware devices.

We assume that the baseline system is made from 
a simple RV32I processor, a local memory, and LEDs for output.
The processor in this system is a multi cycle design
that executes one instruction through five steps consisting of 
instruction fetch (IF), operand fetch (OF), execution (EX), 
memory access (MEM), and write back (WB).

In the IF step, 
an instruction is read from the instruction memory using the program counter (PC) as an address.
In the OF step, 
the register numbers and immediate values are decoded from the instruction and 
the operands are read from the register file.
In the EX step, 
the operation is executed for add, sub, branch instructions and so on.
In the MEM step, 
the main memory is accessed for load and store instructions with the memory address 
which is calculated in the EX step.
In the WB step, 
the value which is calculated in the EX step or read in the MEM step is written back 
to the register file.

We can support M extension for RV32IM smoothly because there are small modifications 
to implement the multiply and divide circuits in the EX step 
if we assume that the multiply and divide can be executed in a single cycle.

%%%%%%%%%%%%%%%%%%%%%%%%%%%%%%%%%%%%%%%%%%%%%%%%%%%%%%%%%%%%%%%%%%%%%%%%%%%%%%%%%%%%%%%%%%%%%%%%%%%%
\subsection{Supporting the control and status register (CSR) instructions}

RISC-V defines some privilege levels where
the execution of some operations can be limited by switching these privilege levels.
There are three privilege levels named {\it M-Mode} for the machine mode,
{\it S-Mode} for the supervisor mode,
and {\it U-Mode} for the user mode~\cite{RISCV-SPEC2}.
The M-Mode is the highest privilege mode and U-Mode is the lowest one.

To identify the privilege mode,
the control and status registers named {\it CSRs} are used.
The registers are also used for exceptions, various identifications
such as supporting instruction extensions, and so on.
To support these registers, it is necessary to implement the CSR instructions,
which operate the read-modify-write on CSRs atomically.

%To implement the CSR instructions,
%%it is necessary to implement an arithmetic and logic unit (ALU) for CSR instructions.

We can implement the CSR instructions by
loading the value of CSRs like the access to the register file in the OF step,
executing in the EX step, and writing the obtained value to CSRs in the WB step.

%%%%%%%%%%%%%%%%%%%%%%%%%%%%%%%%%%%%%%%%%%%%%%%%%%%%%%%%%%%%%%%%%%%%%%%%%%%%%%%%%%%%%%%%%%%%%%%%%%%%
\subsection{Supporting the atomic instructions}

The atomic instructions consist of the load-reserved/store-conditional (LR/SC) instructions 
and the atomic memory operation (AMO) instructions.

The LR instruction loads the data from the main memory and reserves its address.
The SC instruction stores the data to the main memory 
if there is a valid reservation exists on its address.
This instruction also returns a bit that indicates whether the data storing is a success or not.

The complex atomic memory operations can be performed by using LR/SC instructions.
To implement LR/SC instructions, 
it is necessary to implement the registers which save the reservation address 
and the reservation status.
The LR instruction can be implemented 
by writing the reservation address and the status
in the MEM step.
The SC instruction can be implemented by checking the reservation address and status. 
They are small modifications to implement LR/SC instructions on the baseline computer system.

The AMO instructions perform the memory read-modify-write operations.
This means that an AMO instruction atomically loads the value, 
applied the operation, and writes the result to the memory.
It is difficult to implement the AMO instruction on the baseline computer system easily 
because the processor has only one memory access step in the MEM step.
To solve this problem, 
it is necessary to implement an additional memory access step and another execution step 
for the modify operation between these two memory accesses.
These should require a lot of design changes and debugging.

%%%%%%%%%%%%%%%%%%%%%%%%%%%%%%%%%%%%%%%%%%%%%%%%%%%%%%%%%%%%%%%%%%%%%%%%%%%%%%%%%%%%%%%%%%%%%%%%%%%%
\subsection{Supporting the virtual address space}

To support the virtual address space of Linux,
the address translation unit which translates the virtual address to the physical address 
is necessary.

Because each process has its unique virtual address space, 
the virtual address for the space must be translated to the physical address 
to access the physical memory.
Actually, the virtual addresses are managed in units of pages, 
and the translation from these virtual page addresses to the physical page addresses is performed.
We call this translation from virtual address to physical address as {\it page walk}, 
and a translation lookaside buffer (TLB) is a cache that stores the recently used information 
of this translation.

RISC-V defines that the address translation should be done by hardware.
The TLB is accessed before the memory access when the virtual address is used.
If the TLB misses, the page walk is done by hardware.
Furthermore, if the page walk is failed, the page fault exception occurs 
to inform the event to the OS. 
Also, each page has permissions for read, write, and execution.
If there is no valid permission, the page fault exception occurrs too.

The page walk is defined as {\it Sv32} which has 32-bit address space 
having two memory read accesses at maximum.
So, it is difficult to support the address translation on the baseline computer system.

%%%%%%%%%%%%%%%%%%%%%%%%%%%%%%%%%%%%%%%%%%%%%%%%%%%%%%%%%%%%%%%%%%%%%%%%%%%%%%%%%%%%%%%%%%%%%%%%%%%%
\section{RVSoC: a portable and Linux capable RISC-V computer system on an FPGA}

We design a portable and Linux capable RISC-V computer system on a small FPGA, 
and we named {\it RVSoC} for the system.

%%%%%%%%%%%%%%%%%%%%%%%%%%%%%%%%%%%%%%%%%%%%%%%%%%%%%%%%%%%%%%%%%%%%%%%%%%%%%%%%%%%%%%%%%%%%%%%%%%%%
\begin{figure}[tb]
  \centering
    \includegraphics[clip,width=0.7\linewidth]{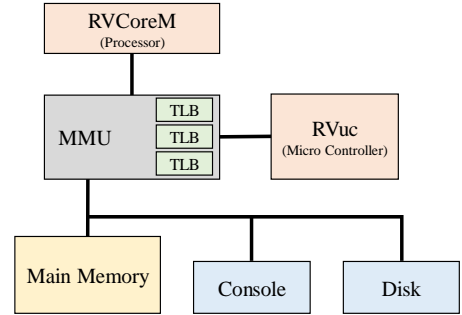}
    \caption{The logical organizaton of the proposed computer system.}
    \label{fig:soc-org}
\end{figure}

Fig.\ref{fig:soc-org} shows the logical organization of RVSoC.
It mainly consists of the main processor named {\it RVCoreM}, 
a memory management unit (MMU), the main memory, a disk, the console, 
and the micro controller named {\it RVuc}.

\RVCoreM \ and \RVuc \ are connected to the main memory, the disk, and the console 
through the MMU.
The MMU has a Sv32 page walk unit to translate the virtual address to the physical address.
It also has three TLBs for the memory access of instruction fetch, data load, and data store, 
respectively.
It is because the pages have permissions such as execution, read, and write.
Therefore, using these three TLBs will simplify the implementation.
The two processors access to the main memory, the disk, and the console 
by using the memory mapped I/O (MMIO).

%%%%%%%%%%%%%%%%%%%%%%%%%%%%%%%%%%%%%%%%%%%%%%%%%%%%%%%%%%%%%%%%%%%%%%%%%%%%%%%%%%%%%%%%%%%%%%%%%%%%
\subsection{RVCoreM: the main processor}

%%%%%%%%%%%%%%%%%%%%%%%%%%%%%%%%%%%%%%%%%%%%%%%%%%%%%%%%%%%%%%%%%%%%%%%%%%%%%%%%%%%%%%%%%%%%%%%%%%%%
\begin{figure*}[tb]
  \centering
    \includegraphics[clip,width=\linewidth]{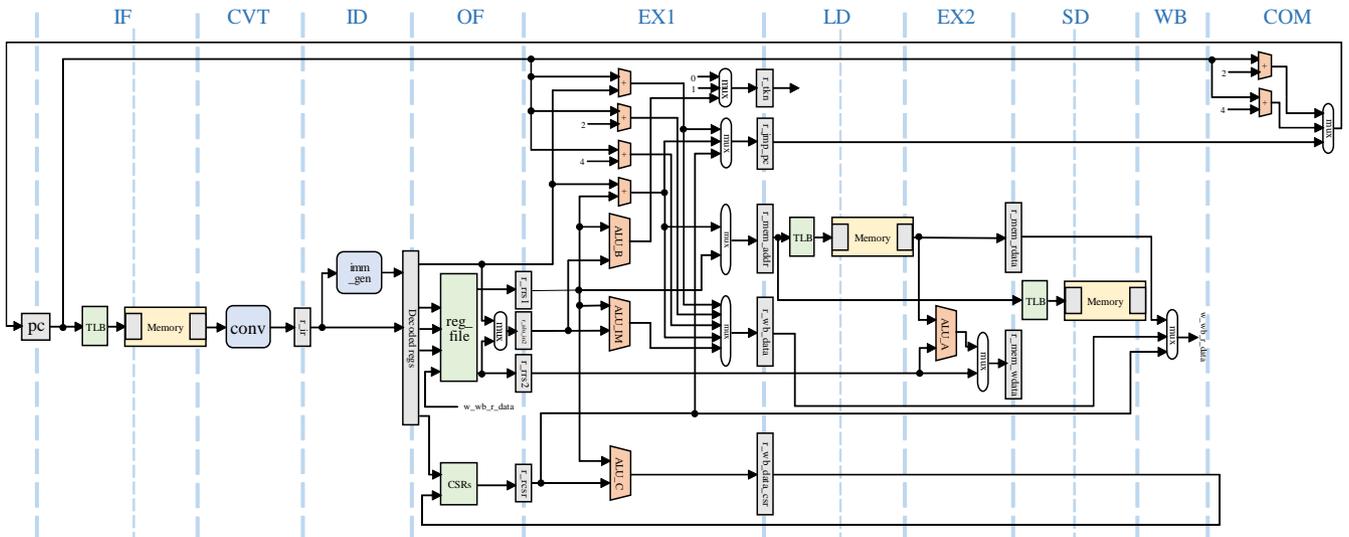}
    \caption{The block diagram of RVCoreM that is the main processor supporting RV32IMAC.}
    \label{fig:rvcorem}
\end{figure*}

Fig.\ref{fig:rvcorem} shows the block diagram of RVCoreM.
It is a multi cycle processor that executes one instruction in 
twelve steps and supports RV32IMAC and CSRs.

The twelve steps are 
initialization (INI), 
IF, 
convert compressed instruction (CVT), 
instruction decode (ID),
OF, execution (EX1), 
load data (LD), 
atomic operate execution (EX2), 
store data (SD), 
WB, update CSR and program counter (COM), 
and increment of the instruction counter (FIN).
INI and FIN are omitted in the figure because they are mainly used 
for the debugging purpose.

In Fig.\ref{fig:rvcorem}, the gray-colored rectangles are registers which are updated at
the positive edge of the clock signal.
The orange-colored units are combinational circuits like ALU and adder.
The yellow-colored unit is the main memory. 
Although three memories are depicted in this figure in the IF, LD, and SD steps, 
they are the same physical memory in fact.
The green-colored units are TLBs, the register file and the CSRs which are
the combinational circuits with asynchronous memories.
The blue-colored units are combinational circuits used in the CVT and ID steps, respectively.

The AMO instructions execute the memory read-modify-write as described above, 
where an AMO instruction loads the data, applies the operation, 
and stores the result to the main memory.
To support the AMO instructions, 
two steps of memory access and the execution step for atomic operation are required.
Therefore, the processor executes an AMO instruction by using the LD step 
for the first memory access, 
the EX2 step for the calculation, and the SD step for the second memory access, 
which perform a memory read-modify-write.

To support the compressed instructions, 
the conversion from a compressed instruction to a standard instruction is done in the CVT step.
The compressed instructions are 16-bit instructions 
that are encoded from standard instructions that meet the specific condition.
Because other standard instructions are 32-bit width, 
the code size can be reduced by using this 16-bit compressed instructions.
Any compressed instruction can be expanded to the equivalent 32-bit standard instruction.
Therefore, 
it is possible to support compressed instructions without adding the complex dedicated circuit 
for compressed instructions in later steps by expanding compressed instructions in the CVT step.

To improve processor performance, we apply two optimizations.
One is the use of multi-cycle divider 
where the divide and the remainder instructions are executed in about 32 cycles 
to improve the operating frequency.
This is because if divide and remainder instructions are executed in one cycle 
like add or sub instructions,
the circuit becomes the critical path of the entire system 
and the operating frequency drops significantly.

The other optimization is that the state transition from IF to FIN is modified
to skip some steps by the operation of the executed instruction to reduce 
the number of elapsed clock cycles for some instructions.
The reason for skipping some steps depending on the executed instruction is 
that even if the instruction does not have memory access, 
the processor has to spend useless cycles in  LD, EX2, and SD steps and this
will degrade performance.

%%%%%%%%%%%%%%%%%%%%%%%%%%%%%%%%%%%%%%%%%%%%%%%%%%%%%%%%%%%%%%%%%%%%%%%%%%%%%%%%%%%%%%%%%%%%%%%%%%%%
\begin{figure}[tb]
  \centering
    \includegraphics[clip,width=0.8\linewidth]{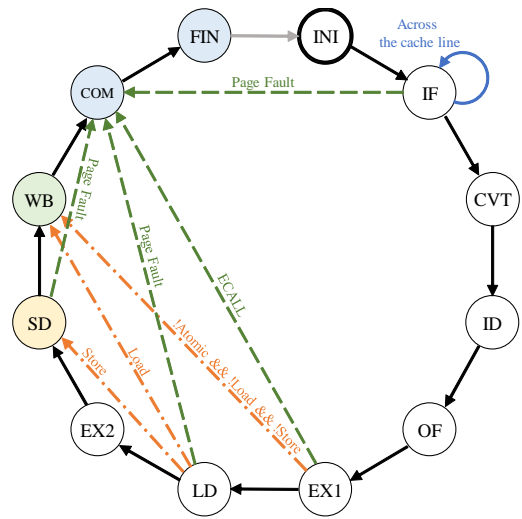}
    \caption{The optimized state transition diagram for the multi cycle design.
The orange, green, and blue transitions are high priority. When these conditions are false, 
the black-colored transitions are used.}
    \label{fig:core-state}
\end{figure}

Fig.\ref{fig:core-state} shows the state transition diagram.
The orange, green, and blue transitions are high priority. When these conditions are false, 
the transition of the black-colored arrow is applied.
All instructions begin from the INI step.

In the IF step, an instruction is fetched. 
To fetch an instruction crossing the cache lines due to the compressed instructions,
the instruction fetch may require two main memory accesses.
If no page fault occurs in the IF step, the next step will be the CVT step.
From the CVT step to the EX1 step are necessary for such instructions.

In the EX1 step, the transition to the WB step for instructions other than 
the atomic, load, and store instructions.
This is because it is sufficient to write back to the register file without accessing the main memory 
expect for the atomic, load, and store instructions.

In the LD step, the load and store instructions do not need to execute atomic operation.
Therefore, the transition to the WB step and SD step, respectively.
By these transitions, 
the average number of execution cycles of the processor can be reduced compared to the 
case of always executing twelve steps (the orange chain line in the fig.\ref{fig:core-state}).

The transition of the green arrow in the Fig.\ref{fig:core-state} 
is for requesting exception handlings.
The exceptions in this figure are the page fault exception and the exception that occurred 
by the ECALL instruction.
When these exceptions occurred, 
the system registers are updated appropriately and the next instruction is executed, 
so that the state transits to the COM step.
Although it is omitted,
when no exception is detected and a stall signal is detected, 
the processor stalls in the current step.

%%%%%%%%%%%%%%%%%%%%%%%%%%%%%%%%%%%%%%%%%%%%%%%%%%%%%%%%%%%%%%%%%%%%%%%%%%%%%%%%%%%%%%%%%%%%%%%%%%%%
\subsection{Translation from the virtual to the physical address}

As described above, 
each page has the execute, read and write access permissions, and
three TLBs for each permission are used.
When there is no permission from the state to be accessed, the TLB is not hit.
When the current step is the IF, the TLB depicted in the IF step in Fig.\ref{fig:rvcorem} 
is accessed.
Similarly,
when it is the LD and SD, the TLB depicted in the LD and SD is accessed, respectively.

When a TLB miss occurs, a page walk is invoked to obtain the translation.
The page walk is designed as a state machine of six states.
This is because it is difficult to access the memory multiple times in one cycle, 
and it is necessary to change the address to access the memory and calculate the 
next address for each state.

In the first state of the page walk, the address of the first translation information is calculated,
and the main memory is accessed with the address.
In the second state, the obtained information is saved in the register.
In the third state, the address of the next translation information is calculated,
and the main memory is accessed with the address again.
In the fourth state, the obtained information is saved in the register.
In the fifth state, the success of the page walk is judged from the saved registers.
In the sixth state, the TLB and the page table entry in the main memory is updated.
These states enable a page walk of RISC-V Sv32.

%%%%%%%%%%%%%%%%%%%%%%%%%%%%%%%%%%%%%%%%%%%%%%%%%%%%%%%%%%%%%%%%%%%%%%%%%%%%%%%%%%%%%%%%%%%%%%%%%%%%
\subsection{I/O devices and its controller}

%%%%%%%%%%%%%%%%%%%%%%%%%%%%%%%%%%%%%%%%%%%%%%%%%%%%%%%%%%%%%%%%%%%%%%%%%%%%%%%%%%%%%%%%%%%%%%%%%%%%
\begin{figure}[tb]
  \centering
    \includegraphics[clip,width=\linewidth]{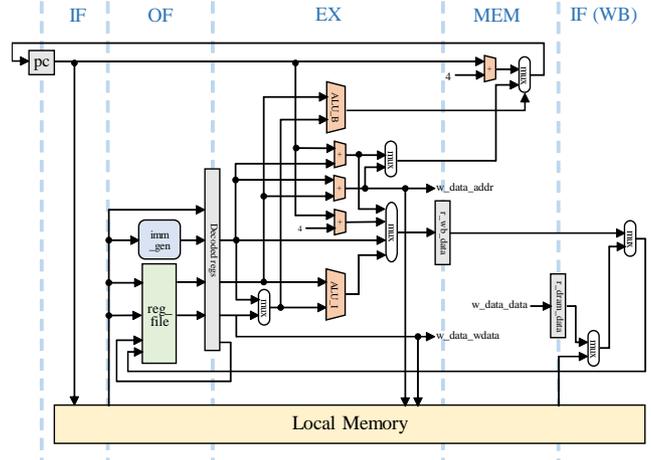}
    \caption{The block diagram of the I/O controller named RVuc.}
    \label{fig:rvuc}
\end{figure}

As shown in Fig.\ref{fig:soc-org}, we implement the console and the disk as I/O devices.
The console is a device for a keyboard input and a character output,
and the disk is a storage device having a file system.

These devices are accessed using VirtIO~\cite{virtio} that is an I/O framework.
The VirtIO is generally used by the emulator, 
and when accessing the actual I/O, many main memory accesses and loops depending 
on the contents of the memory data occur.
If this operation is implemented as hardware, a complex circuit is required.
Therefore, 
a small processor named RVuc is implemented and used as an I/O controller 
to execute these complicated processes as software.

Fig.\ref{fig:rvuc} shows the block diagram of \RVuc .
It is a four step and multi cycle processor having own local memory which supports RV32I.

The programs of the I/O processing for the console and the disk are stored in this local memory.
RVuc can access to DRAM or I/O control registers 
by using w\_data\_addr, w\_data\_wdata, and  w\_data\_data wires in Fig.\ref{fig:rvuc}.

\RVuc \ executes the program and I/O processing is done when the request of I/O is  
came from \RVCoreM .
RVCoreM stops the operation while \RVuc \ operating.
There is no additional TLB and control registers for RVuc
because RVuc does not use the virtual address space.

%%%%%%%%%%%%%%%%%%%%%%%%%%%%%%%%%%%%%%%%%%%%%%%%%%%%%%%%%%%%%%%%%%%%%%%%%%%%%%%%%%%%%%%%%%%%%%%%%%%%
\subsection{Memory access optimization} %% Cache memory}

A direct mapped cache of the write through scheme is implemented 
to reduce the access latency to the main memory, and
the memory access latency is one cycle when the cache hits.
The cache block size is 16-byte.
For a simple implementation, the cache entry is invalidated 
when a corresponding entry is updated by a store instruction.

This cache is located between the main memory and MMU.
Because all the memory addresses to the cache are physical ones, 
the cache flush due to the process switching is not occurred. 
Note that this cache works as an instruction cache and a data cache for both RVCoreM and RVuc.

The instruction fetch unit always requests 4-byte data to the cache.
However, 4-byte fetches may be fetched across cache lines when supporting compressed instructions.

%%%%%%%%%%%%%%%%%%%%%%%%%%%%%%%%%%%%%%%%%%%%%%%%%%%%%%%%%%%%%%%%%%%%%%%%%%%%%%%%%%%%%%%%%%%%%%%%%%%%
\begin{figure}[tb]
  \centering
    \includegraphics[clip,width=0.98\linewidth]{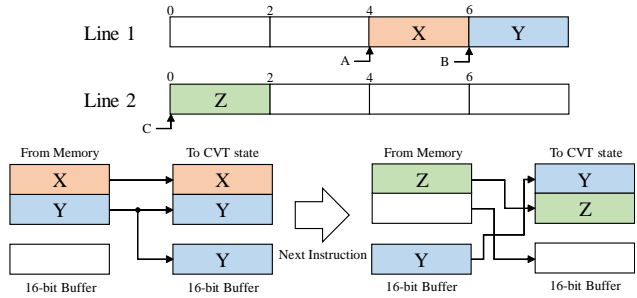}
    \caption{The instruction fetching across two cache lines and the scheme using the 16-bit Buffer.}
    \label{fig:16bitBuffer}
\end{figure}

Fig.\ref{fig:16bitBuffer} shows the way to fetch 4-byte data across two cache lines.
For simplicity, we assume that the line size is 8-byte, 
and Line1 and Line2 store the continuous blocks.
If the value of the program counter is at the position A in the Fig.\ref{fig:16bitBuffer},
the 4-byte of X and Y which are 2-bytes each can be fetched since it does not cross cache lines.
If the value of the program counter is at the position B, Y and Z have to be fetched.
At this time, two cache lines have to be accessed since Y and Z are on different cache lines.
These two accesses increase the number of cycles by processor stall and decrease
the performance.

We mitigate this problem by implementing a small buffer named {\it 16-bit Buffer}.
The 16-bit Buffer stores the upper 2-byte of the previously fetched 4-byte instruction.
In Fig.\ref{fig:16bitBuffer}, the value of the program counter is at the position A, 
and when X and Y are fetched, and Y is stored in the 16-bit Buffer.

Then, when the value of the program counter is at the position of B for the next instruction, 
it fetches 4-byte including Z.
The fetched instruction is completed by concatenating the fetched Z and Y in the 16-bit Buffer 
and sending it to the CVT step.
This operation can reduce the number of cache accesses when fetching 4-bytes across the cache line.

%%%%%%%%%%%%%%%%%%%%%%%%%%%%%%%%%%%%%%%%%%%%%%%%%%%%%%%%%%%%%%%%%%%%%%%%%%%%%%%%%%%%%%%%%%%%%%%%%%%%
\subsection{Implementation targetting Nexys A7 FPGA board} %% of the computer system RVSoC}

We describe the implementation issues on RVSoC 
in Verilog HDL targetting Nexys A7 FPGA board of Digilent Inc.

%%%%%%%%%%%%%%%%%%%%%%%%%%%%%%%%%%%%%%%%%%%%%%%%%%%%%%%%%%%%%%%%%%%%%%%%%%%%%%%%%%%%%%%%%%%%%%%%%%%%
\begin{figure}[tb]
  \centering
    \includegraphics[clip,width=0.8\linewidth]{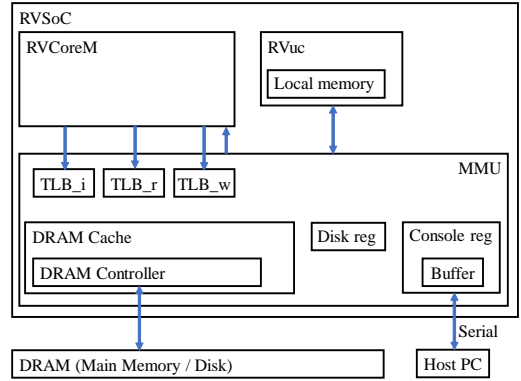}
    \caption{The organization of RVSoC implemented in Verilog HDL targetting Nexys A7 FPGA board.}
    \label{fig:verilog-org}
\end{figure}

Fig.\ref{fig:verilog-org} shows the hardware organization implemented in Verilog HDL.
The DRAM memory and the host PC are outside of an FPGA.
TLB\_i, TLB\_r, and TLB\_w are modules of TLBs for instruction, read, and write, respectively.
There is no TLB between RVuc and DRAM cache module 
because RVuc uses physical addresses in the program execution.
The access to the DRAM memory, the console registers, and the disk registers is controlled
by the MMIO module.

The Nexys A7 board has 128MB DDR2-SDRAM.
All reads to the DRAM are performed in 16-byte units because the cache line size is 16-byte.
The writes are performed in 1-byte, 2-byte, or 4-byte units 
executing SB (store byte), SH (store halfword), 
and SW (store word), respectively.
Accessing to this DRAM uses Memory Interface Generator (MIG), an IP of Xilinx, Inc.
The operation frequency of the DRAM is 325MHz, 
which is the maximum operating frequency recommended by the Nexys A7 manual~\cite{nexysa7},
and the operating frequency of the DRAM controller is 81.25MHz which is 1/4 of the DRAM frequency.

The 64MB DRAM memory area is used for the main memory, 
and the rest of the 64MB area is used for the disk.
Therefore, the system does not use any physical disk drive.

For the console input and output, serial communication is used.
A FIFO buffer for 16 characters for keyboard input is implemented to prevent 
the omission of detection due to high-speed input.
The serial communication is also used 
to initiate the contents of the main memory and the disk.
The communication speed of the serial communication between an FPGA and the host PC is 8Mbaud.

%%%%%%%%%%%%%%%%%%%%%%%%%%%%%%%%%%%%%%%%%%%%%%%%%%%%%%%%%%%%%%%%%%%%%%%%%%%%%%%%%%%%%%%%%%%%%%%%%%%%
\begin{figure}[tb]
  \centering
    \includegraphics[clip,width=\linewidth]{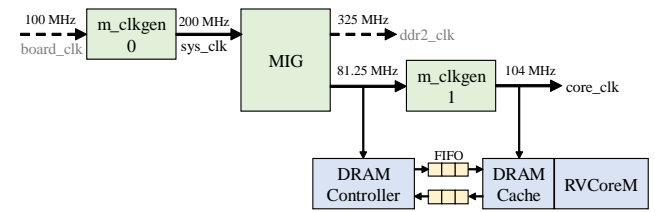}
    \caption{The clock generation scheme on an FPGA using the 100MHz input clock.}
    \label{fig:clock}
\end{figure}

Fig.\ref{fig:clock} shows the clock generation scheme on an FPGA using the 100MHz input clock.
The dashed lines indicate input and output clocks to and from an FPGA.

The 200MHz clock is generated from the 100MHz input clock 
using the module m\_clkgen0, which is the input clock of the module MIG.
Using this 200MHz clock, the 325MHz clock for the DRAM memory 
and the 81.25MHz clock for the DRAM controller are generated.
Then, using this 81.25MHz clock as input, 
the module m\_clkgen1 generates the 104MHz clock for RVCoreM and RVuc.
Since the operating frequency of the memory controller running at 81.25MHz and
the frequency of processors running at 104MHz are different,
the data transfer between them is done using asynchronous FIFOs.

%%%%%%%%%%%%%%%%%%%%%%%%%%%%%%%%%%%%%%%%%%%%%%%%%%%%%%%%%%%%%%%%%%%%%%%%%%%%%%%%%%%%%%%%%%%%%%%%%%%%
\begin{table}[tb]
  \centering
  \caption{The number of lines of Verilog HDL code for each file and their total.}
  \begin{tabular}{lr|lr}
    \hline
    console.v & 119 & memory.v    & 841  \\
    debug.v   & 161 & microc.v    & 239  \\
    disk.v    & 118 & mmu.v       & 798  \\
    dram.v    & 515 & rvcorem.v   & 1,263 \\
    loader.v  & 143 & top.v       & 557  \\ \cline{3-4} 
    main.v    & 359 & total       & 5,113 \\ \hline
    \end{tabular}
    \label{tab:Verilog}
\end{table}

Table \ref{tab:Verilog} lists the names of all RTL design files and 
shows the number of lines of Verilog HDL code 
for each file and their total.
Note that some necessary header files are not listed in this table.
 The file of main.v is the top module for connecting RVCoreM to the MMU and 
for controlling LEDs output on the FPGA.
rvcorem.v, microc.v, and mmu.v are files of RVCoreM, RVuc, and MMU, respectively.
dram.v is a file for DRAM controller to use the DRAM,
memory.v is a file that defines the cache and local memory of RVuc,
and loader.v is a file for receiving the initialization file 
and sending and receiving the terminal,
and console.v and disk.v are files controlling each system register,
and debug.v is a file for displaying debug information on the terminal.

The total number of simulation files is 5,113 lines.
The file top.v is used for only simulation, 
the number of Verilog HDL code lines for the FPGA logic synthesis excluding top.v is 4,556.

%%%%%%%%%%%%%%%%%%%%%%%%%%%%%%%%%%%%%%%%%%%%%%%%%%%%%%%%%%%%%%%%%%%%%%%%%%%%%%%%%%%%%%%%%%%%%%%%%%%%
\section{Verification and evaluation}

%%%%%%%%%%%%%%%%%%%%%%%%%%%%%%%%%%%%%%%%%%%%%%%%%%%%%%%%%%%%%%%%%%%%%%%%%%%%%%%%%%%%%%%%%%%%%%%%%%%%
\subsection{Verification and evaluation environment}

We describe the evaluation environment to evaluate the hardware resources and performance.
The used FPGA board is Digilent Nexys A7 with xc7a100tcsg324-1 FPGA and 128MB DDR2-SDRAM.

Vivado Design Suite 2017.2 from Xilinx, Inc 
is used for the synthesis and implementation for an FPGA.
The Flow\_PerfThresholdCarry (Vivado Synthesis 2017) strategy is used for synthesis, 
and the Flow\_RunPostRoutePhysOpt (Vivado Implementation 2017) strategy is used for implementation.

%%%%%%%%%%%%%%%%%%%%%%%%%%%%%%%%%%%%%%%%%%%%%%%%%%%%%%%%%%%%%%%%%%%%%%%%%%%%%%%%%%%%%%%%%%%%%%%%%%%%
\begin{table}[tb]
  \centering
  \caption{The main parameters for the evaluation.}
  \begin{tabular}{l|rl}
    \hline
    \RVSoC \ Version       & \RVver & \\
    Core clock frequency   & 104 & MHz \\
    DRAM clock frequency   & 325 & MHz \\
    TLB entries            & 32$\times$3 & Entries \\
    Cache size             & 128 & KB   \\
    Cache data width       & 16  & Byte \\
    RVuc local memory size & 8   & KB   \\ \hline
    \end{tabular}
    \label{tab:parameter}
\end{table}

Table \ref{tab:parameter} shows the main parameters for the evaluation, 
and we use this parameter unless otherwise noted.
The core clock frequency for two processors and MMU is set to 104MHz.
The number of entries for each TLB is 32, and the direct mapped scheme is used for three TLBs.

%% The number of lines of Verilog HDL code for the implementation of RVSoC is 5,113 lines.

We use TeraTerm on the host PC running Windows 10 to send the initialization data to the console.
The baud rate is set to 8Mbaud, 
and the data of the boot loader and the disk image is transmitted after 
writing the bit stream to the FPGA.

The Linux kernel to be executed is version 4.15.0, 
and the root file was built for two configurations using Buildroot~\cite{Buildroot}.
The one is targetting RV32IMAC (with the compressed instructions).
The other is targetting RV32IMA (without the compressed instructions).

%%The host PC and Nexys A7 are connected via a USB cable.

%%%%%%%%%%%%%%%%%%%%%%%%%%%%%%%%%%%%%%%%%%%%%%%%%%%%%%%%%%%%%%%%%%%%%%%%%%%%%%%%%%%%%%%%%%%%%%%%%%%%
\subsection{Verification}

We verified RVSoC by using Verilog simulation, a software simulator in C++, 
and an FPGA board.
As a software simulator, 
we use {\it SimRV}~\cite{SimRV} which can emulate an RISC-V computer system.

For each executed instruction, 
SimRV can output the architectural state which includes the contents of the program counter, 
the instruction register, the general purpose registers, the CSRs, and the TLBs.
Similarly, we implemented the function to output the same information with the same format
as SimRV to the RTL of RVSoC.

%% However, in SimRV, the mtime register, which is a counter that is updated at regular intervals, 
%% is counted as the number of instructions to be executed.
%% Therefore, RVSoC should be designed accordingly.
%% This means the actual time does not match the time in RVSoC correctly.

For verification, the architectural states obtained from SimRV and 
the Verilog simulation of RVSoC 
were compared.
Synopsys VCS is used for simulation to obtain the architecture state of RVSoC.
From the comparison, 
we confirmed that both architectural states for all simulated instructions match completely.

%%%%%%%%%%%%%%%%%%%%%%%%%%%%%%%%%%%%%%%%%%%%%%%%%%%%%%%%%%%%%%%%%%%%%%%%%%%%%%%%%%%%%%%%%%%%%%%%%%%%
\subsection{Evaluation}

%%%%%%%%%%%%%%%%%%%%%%%%%%%%%%%%%%%%%%%%%%%%%%%%%%%%%%%%%%%%%%%%%%%%%%%%%%%%%%%%%%%%%%%%%%%%%%%%%%%%
\begin{figure}[tb]
  \centering
    \includegraphics[clip,width=0.9\linewidth]{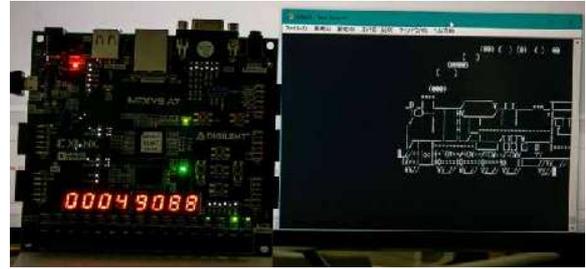}
    \caption{The photo of RVSoC on Nexys A7 FPGA board (left) 
and a screenshot execution {\it sl} command on TeraTerm (right).}
    \label{fig:pic}
\end{figure}

Fig.\ref{fig:pic} is 
the photo of RVSoC working on Nexys A7 FPGA board 
and a screenshot on TeraTerm (right).
The left side of the Fig.\ref{fig:pic} is RVSoC working on Nexys A7 FPGA board.
The number displayed on the 7-segment LED on the FPGA board is a hexadecimal 
value about the number of executed instructions.
%obtained by dividing the number of executed instructions by 1,024.
The right side of the figure is a screenshot execution {\it sl} command 
on TeraTerm displayed by serial communication.
Besides the sl command, various commands on Linux 
such as {\it top},{\ sleep}, and {\it vi} can be executed.

%%%%%%%%%%%%%%%%%%%%%%%%%%%%%%%%%%%%%%%%%%%%%%%%%%%%%%%%%%%%%%%%%%%%%%%%%%%%%%%%%%%%%%%%%%%%%%%%%%%%
\begin{figure}[tb]
  \centering
    \includegraphics[clip,width=0.9\linewidth]{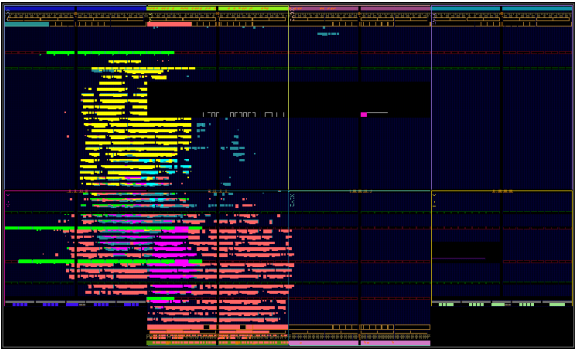}
    \caption{The implementation result of \RVSoC \ on a xc7a100tcsg324-1 FPGA.}
    \label{fig:impl}
\end{figure}

Fig.\ref{fig:impl} shows the implementation result of \RVSoC \ on a xc7a100tcsg324-1 FPGA.
In this figure, 
the yellow blocks are RVCore, the purple blocks are RVuc, the orange blocks are the DRAM controller, 
the green blocks are the cache, and the light blue blocks are TLBs.

%%%%%%%%%%%%%%%%%%%%%%%%%%%%%%%%%%%%%%%%%%%%%%%%%%%%%%%%%%%%%%%%%%%%%%%%%%%%%%%%%%%%%%%%%%%%%%%%%%%%
\begin{table}[tb]
  \centering
  \caption{The hardware resources.}
  \begin{tabular}{c|c|c}
    \hline
    Registers     & LUTs            & BRAMs       \\ \hline
    6,379 (5.0\%) & 10,421 (16.4\%) & 38 (28.2\%) \\ \hline
  \end{tabular}
  \label{tab:eval-impl}
\end{table}

Table \ref{tab:eval-impl} shows the number of the occupied hardware resources of RVSoC
where the number in parentheses indicate the percentage of the whole FPGA resources.
When targeting the xc7a100tcsg324-1 FPGA, 
the utilization rate of hardware resources is less than 30\%, 
and there is much space to implement additional logic such as accelerators.

%%%%%%%%%%%%%%%%%%%%%%%%%%%%%%%%%%%%%%%%%%%%%%%%%%%%%%%%%%%%%%%%%%%%%%%%%%%%%%%%%%%%%%%%%%%%%%%%%%%%
\begin{table}[tb]
  \centering
  \caption{The cycles per instruction (CPI) on two configurations.}
  \begin{tabular}{l|rr|r}
  \hline
           & Executed insns  & Cycles        & CPI  \\ \hline
   Comp    & 66,067,456      & 1,213,305,856 & 18.4 \\ 
   No-comp & 66,760,704      & 1,233,961,984 & 18.5 \\ \hline
   \end{tabular}
   \label{tab:ipc-eval}
\end{table}

Table \ref{tab:ipc-eval} shows the cycles per instruction (CPI) of RVCoreM
measured until the login screen of Linux is displayed on the console.
The configuration of {\it Comp} is that all programs including the Linux kernel are
compiled targetting RV32IMAC with compressed instructions.
The configuration of {\it No-comp} is that all programs are
compiled targetting RV32IMA without compressed instructions.

The second and third column in this table indicate the number of executed instructions
and elapsed cycle, respectively.
The number of elapsed cycles is not counted while RVuc is executing the I/O processing 
and RVCoreM is stalling the operation.

Although the minimum number of cycles to execute one instruction is eight, 
the average numbers of the execution cycles per instruction are 18.4 and 18.5 
because the DRAM memory accesses take many clock cycles.

%%%%%%%%%%%%%%%%%%%%%%%%%%%%%%%%%%%%%%%%%%%%%%%%%%%%%%%%%%%%%%%%%%%%%%%%%%%%%%%%%%%%%%%%%%%%%%%%%%%%
\begin{table}[tb]
  \centering
  \caption{The cache hit rate and the miss per kilo instruction (MPKI).}
  \begin{tabular}{l|rr|rr}
  \hline
          & Access Num   & Hit Num     & Hit rate (\%) & MPKI  \\ \hline
  Comp    & 86,738,837   & 82,231,973  & 94.8          & 73.88 \\
  No-comp & 77,055,765   & 71,725,363  & 93.1          & 87.38 \\ \hline
  \end{tabular}
  \label{tab:cache-hit}
\end{table}

Table \ref{tab:cache-hit} shows the cache hit ratio and miss per kilo instruction (MPKI) 
on two configurations
measured during the 61M instruction was executed 
until the Linux login screen was displayed in the simulation.
We can see that the hit rates of the cache are more than 93\% for both configurations,
and the cache is really effective for the implementation of computer systems.

%%%%%%%%%%%%%%%%%%%%%%%%%%%%%%%%%%%%%%%%%%%%%%%%%%%%%%%%%%%%%%%%%%%%%%%%%%%%%%%%%%%%%%%%%%%%%%%%%%%%
\begin{figure}[tb]
  \centering
    \includegraphics[clip,width=\linewidth]{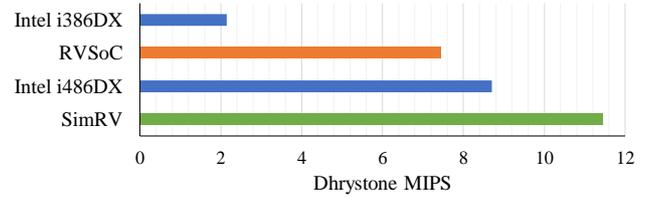}
    \caption{The performance comparison of RVSoC and some computer systems.}
    \label{fig:dmips}
\end{figure}

Fig.\ref{fig:dmips} shows the performance comparison of RVSoC and some computer systems
by using Dhrystone MIPS (DMIPS) as the metric
when Dhrystone benchmark~\cite{Dhrystone} of Buildroot package is executed.
The evaluation targets are Intel i386DX~\cite{microcontrollers}, 
Intel i486DX~\cite{microcontrollers}, 
SimRV\footnote{Running on Intel Core i7 870, DDR3-SDRAM 8GB, Ubuntu 16.04 machine}, 
and RVSoC.
Our proposal of RVSoC achieves 
almost the same performance as the Intel Core i486DX that is a processor about 30 years ago.
Unfortunately, the software simulator SimRV achieves higher performance than 
RVSoC running on an FPGA, but we see that there is no significant difference.

%%%%%%%%%%%%%%%%%%%%%%%%%%%%%%%%%%%%%%%%%%%%%%%%%%%%%%%%%%%%%%%%%%%%%%%%%%%%%%%%%%%%%%%%%%%%%%%%%%%%
\begin{table}[tb]
  \centering
  \caption{The time from the synthesis using Vivado to the display of the Linux login prompt.}
  \begin{tabular}{rrrr|r}
    \hline
    Synthesis  & Imple \& bit gen  & Initialize  &  Boot   & Total   \\ \hline
    124 sec    & 161 sec       & 33 sec      &  12 sec & 330 sec \\ \hline
  \end{tabular}
  \label{tab:impl-time}
\end{table}

Table \ref{tab:impl-time} shows
the time from the synthesis using Vivado to the display of the Linux login prompt.
on a Nexys A7 FPGA board.
The synthesis and implementation are running on Intel Core i9 7920X with 64GB DDR4-SDRAM 
running Ubuntu 18.04.
The five columns indicate the time for
(1) the synthesis, 
(2) the place, route and a bitfile generation,
(3) the transferring the boot loader image of 9.1MB and the disk image of 16MB,
(4) the Linux booting, and
(5) their total.

Although the maximum size of the disk is 64MB, only 16MB image is used because 
the larger data transfer takes much initialization time.
This system is easy-to-use because it can be executed in less than 6 minutes in total 
including synthesis, implementation, and bit file generation time.

%%%%%%%%%%%%%%%%%%%%%%%%%%%%%%%%%%%%%%%%%%%%%%%%%%%%%%%%%%%%%%%%%%%%%%%%%%%%%%%%%%%%%%%%%%%%%%%%%%%%
\section{Related works and discussion}

%%%%%%%%%%%%%%%%%%%%%%%%%%%%%%%%%%%%%%%%%%%%%%%%%%%%%%%%%%%%%%%%%%%%%%%%%%%%%%%%%%%%%%%%%%%%%%%%%%%%
\subsection{Related works}

Freedom U500 VC707 Dev Kit platform of SiFive is an SoC that supports Linux, and RV64GC (RV64IMAFDC).
This system is built from Chisel~\cite{Chisel} source files.
The computer system can utilize from one to eight RV64GC processor cores, 
each with its own private L1 cache and shared L2 cache, and DDR3/4 DRAM as well as 1Gb Ethernet~\cite{U500}.
It also supports custom instruction extension and accelerators.
The high performance processor core has a single-issue, in-order 64-bit execution pipeline.
This computer system is available for FPGAs using the Xilinx VC707.
VC707 is an extremely high performance embedded systems and very expensive (\$3,495~\cite{VC707}).
Therefore, it is not portable and is hard to use for those 
who are starting to develop RISC-V software.

Litex-VexRiscv~\cite{VexSoc} is a computer system equipped with a processor 
called VexRiscv~\cite{VexRiscv}.
VexRiscv is a processor can support Linux and RV32IMA
and written in SpinalHDL~\cite{SpinalHDL} which is the original HDL based on Scala.
An instruction cache and a data cache are implemented.
This computer system can be implemented on many FPGA boards by using FPGA design / SoC 
builder called Litex~\cite{Litex}.
VexRiscv is written in SpinalHDL 
and it is not easy-to-use 
because Verilog HDL and SystemVerilog are dominant languages used to implement 
processors~\cite{Survey-RISCV}.

%%%%%%%%%%%%%%%%%%%%%%%%%%%%%%%%%%%%%%%%%%%%%%%%%%%%%%%%%%%%%%%%%%%%%%%%%%%%%%%%%%%%%%%%%%%%%%%%%%%%
\subsection{Obtained knowledge from the development}

The development period of RVSoC was about half a year, through the development we obtained various findings
as follows.
%The main obtained knowledges are
(1) the improvement of the debugging efficiency by using a high-speed simulator in C++ on the system design,
(2) the importance of the output function of the architecture state,
(3) the importance of the function to restart the simulation from any point,
and (4) the benefits of performing the complex processings with a small processor.
We will explain these each by each.

We used SimRV which is written in C++ to design RVSoc.
Because Verilog simulation is much slower than C++,
using a high-speed software simulator such as SimRV makes it easier to design the processor.
SimRV has a hardware-like design,
so design changes of the system can apply to Verilog HDL code easily.
As a result, it was possible to design the RVSoC more quickly 
than when designing it using only Verilog HDL code.

The output function of the architecture state is used to find implementation differences 
between software simulators and the hardware design by Verilog HDL.
it is possible to easily identify the points where the difference occurred by adding this output function.
This has made it easier to find and fix bugs in the hardware design quickly.

The function that can restart the simulation by Verilog HDL from the middle is necessary
because the Verilog HDL simulation is slower than the software simulator.
The target of the implementation for operating Linux in the simulation
was up to the 61Mth instruction to display the Linux login screen,
but it would take more than 30 minutes and the log data would be enormous
if this simulation was executed by Synopsys VCS.
This means that you will have to wait a long time when debugging the operation in the latter half of the simulation,
Therefore, the architecture state, and the contents of the memory and the disk 
are all saved in one log file at the end of simulation,
and it was possible to simulate only the necessary parts from the time when the log file was acquired.
As a result, this restart function contributed to shortening the development period.

The complex processing is executed by a micro controller which is a small processor.
The complex I/O processing is used in software simulators.
If the implementation of I/O processing in hardware is different from the processing used 
in the software simulators,
the architectural state will not match the software simulator because different I/O processings are executed.
This makes debugging including I/O processing difficult.
Therefore,
not only the system was simplified, but also debugging was easy
by implementing a micro controller that performs exactly the same processing as software simulation.

%%%%%%%%%%%%%%%%%%%%%%%%%%%%%%%%%%%%%%%%%%%%%%%%%%%%%%%%%%%%%%%%%%%%%%%%%%%%%%%%%%%%%%%%%%%%%%%%%%%%
\subsection{Expected usage of \RVSoC}

We are planning to release the RTL code of the designed RVSoC as an open and royalty free RTL design.

Because RVSoC is a computer system that supports Linux and 
uses a small amount of hardware resources, 
it can be applied to various purposes.

A feature of RISC-V is that it has a room for the extended instructions 
by computer system developers.
The ability of extension can be the basic requirement for application-specific 
accelerators and it enables to implement more specialized instruction sets.
For example, 
the RISC-V processor core of the PULP Platform~\cite{PULP} has improved performance 
by adding some packed-SIMD instructions, some bit manipulation instructions and so on.
The resource-saving of RVSoC can be suitable for 
the implementation of various accelerators and special processor cores 
by adding unique instructions, and the development of related software.

The number of lines in Verilog HDL code of RVSoC is about 5,000, 
and it is relatively easy to understand the entire implementation 
of the Linux capable computer system.
Therefore, 
it is suitable to be used as a sample computer system of the education 
on computer science.

%%%%%%%%%%%%%%%%%%%%%%%%%%%%%%%%%%%%%%%%%%%%%%%%%%%%%%%%%%%%%%%%%%%%%%%%%%%%%%%%%%%%%%%%%%%%%%%%%%%%
\section{Conclusion}

A Linux capable computer system has to support the CSR instructions, 
the atomic instructions, and the virtual address space.
The atomic instructions are difficult to be implemented on a simple computer 
system because there are instructions that access the main memory twice per instruction.

We proposed RVSoC, a portable and Linux capable RISC-V computer system 
which is implemented in Verilog HDL.
It mainly consists of a processor named RVCoreM,
a memory management unit, the main memory, a disk, the console
and a micro controller named RVuc.
RVCoreM is a twelve step and multi-cycle processor that supports RV32IMAC, 
and supports the atomic instructions by implementing two memory access steps.
RVuc is a small processor used for the disk and console accesses.

The evaluation results show that
RVSoC can be implemented with a small amount of hardware resources such as 
registers of about 5\%, LUTs of about 16\%, 
and BRAMs of about 28\% of the target FPGA.
The RTL code of the system is about 5,000 lines, which makes it easy 
to understand the implementation.
The time from the synthesis using Vivado to the display of the Linux login 
screen is less than 6 minutes. 
Such a short development time makes the proposed system portable and easy-to-use.

%\ack
%%%%%%%%%%%%%%%%%%%%%%%%%%%%%%%%%%%%%%%%%%%%%%%%%%%%%%%%%%%%%%%%%%%%%%%%%%%%%%%%%%%%%%%%%%%%%%%%%%%%
\section*{Acknowledgments}
This work was supported by JSPS KAKENHI Grant Number JP16H02794.
This work is supported by VLSI Design and Education Center(VDEC), 
the University of Tokyo in collaboration with Synopsys, Inc.
We thank Mr. Kuroda for creating SimRV.

%\nocite{}

%%%%%%%%%%%%%%%%%%%%%%%%%%%%%%%%%%%%%%%%%%%%%%%%%%%%%%%%%%%%%%%%%%%%%%%%%%%%%%%%%%%%%%%%%%%%%%%%%%%%
\bibliographystyle{ieicetr}% bib style
\bibliography{myrefs}% your bib database
%\begin{thebibliography}{99}% more than 9 --> 99 / less than 10 --> 9
%\bibitem{}
%\end{thebibliography}

\profile[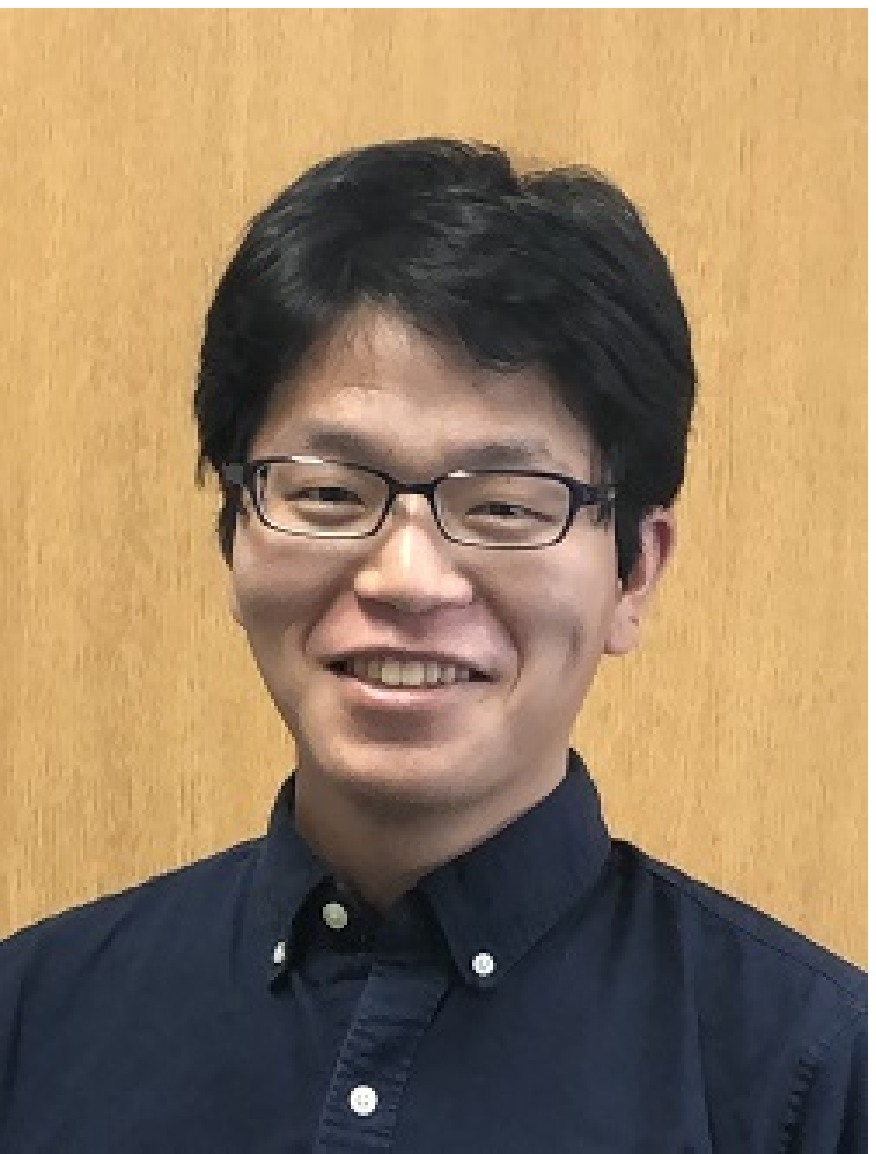]{Junya Miura}{received the B.E degrees in Department of Computer Science
from Tokyo Institute of Technology, Japan in 2018.
He is currently a master course student of the Graduate School of Computing, Tokyo Institute of Technology, Japan.
His research interest is computer architecture, high performance computing and FPGA computing.}
\profile[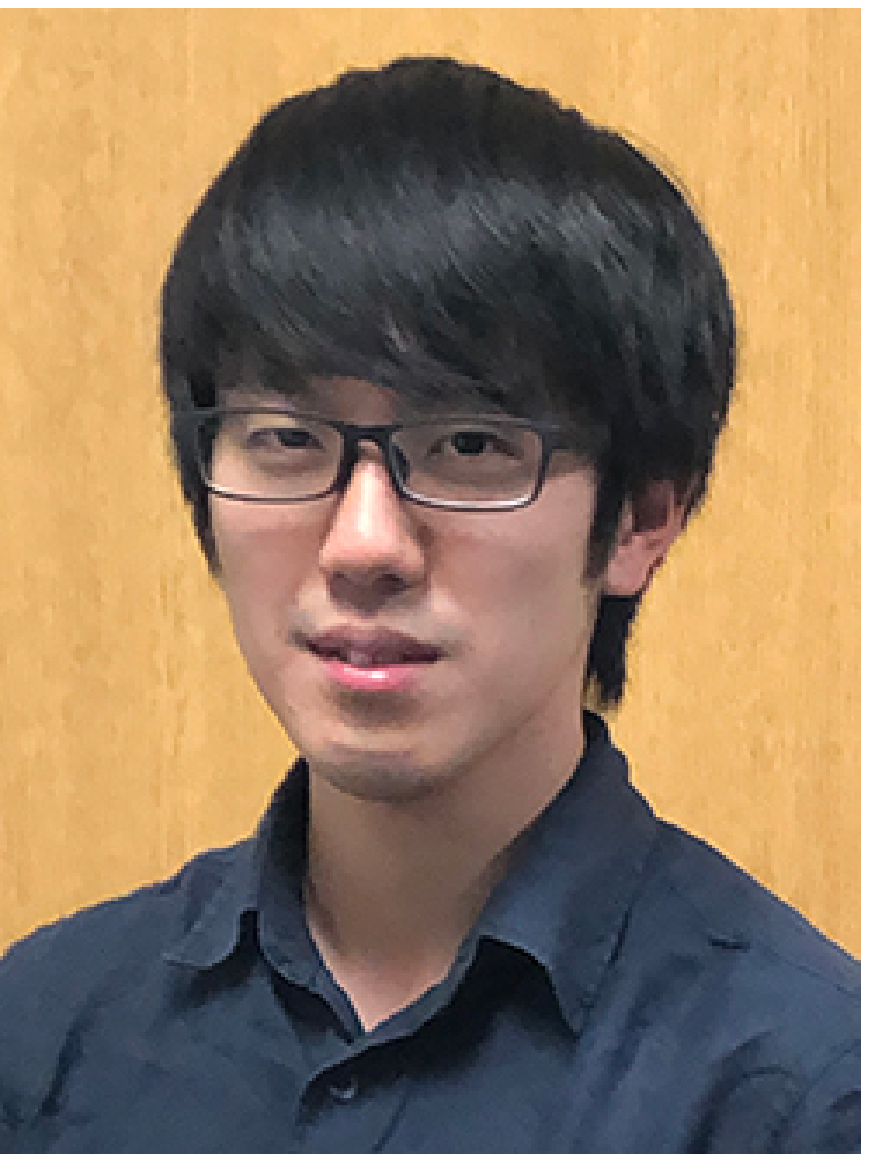]{Hiromu Miyazaki}{received the B.E degrees in Department of Computer Science
from Tokyo Institute of Technology, Japan in 2019.
He is currently a master course student of the Graduate School of Computing, Tokyo Institute of Technology, Japan.
His research interest is computer architecture and FPGA computing.
He is a student member of IEICE.}
\profile[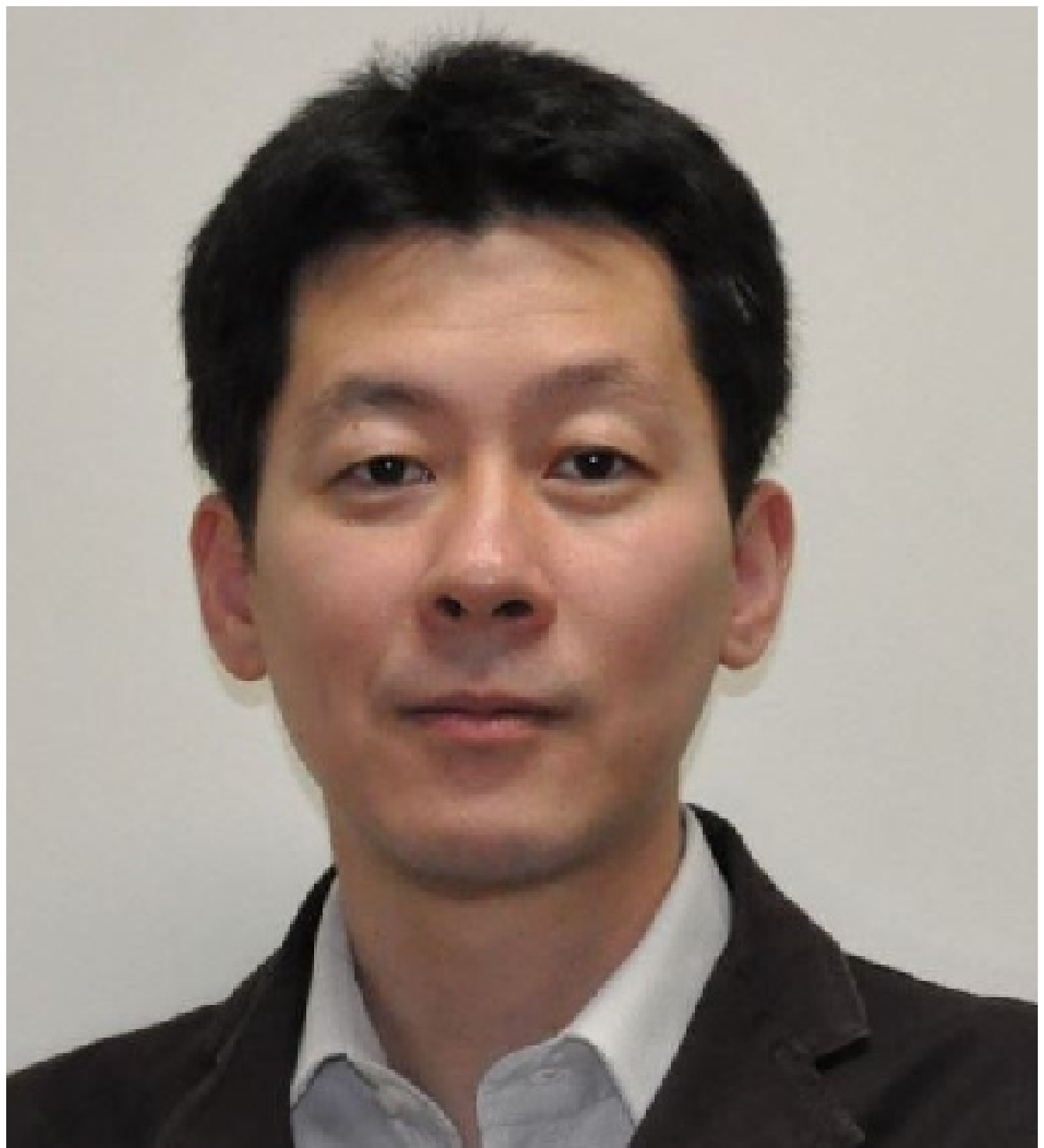]{Kenji Kise}{received the B.E. degree from Nagoya University in 1995, the M.E. degree and the Ph.D. degree in information engineering from the University of Tokyo in 1997 and 2000, respectively.
He is currently an associate professor of the School of Computing, Tokyo Institute of Technology.
His research interests include computer architecture and parallel processing. He is a member of ACM, IEEE, IEICE, and IPSJ.}
%\label{profile}

%\profile{}{}
%\profile*{}{}% without picture of author's face

\end{document}